\documentclass[12pt,preprint]{aastex}

\newcommand{\skipthis}[1]{}
\newcommand{\kms}{\hbox{km\,s$^{-1}$}}

\slugcomment{Accepted by ApJ}
\shorttitle{HCN 4--3 in IRAS 16293-2422}
\shortauthors{Takakuwa et al.}

\begin{document}
\title{Arcsecond-Resolution Submillimeter HCN Imaging of\\
the Binary Protostar IRAS 16293-2422}

\author{Shigehisa Takakuwa\altaffilmark{1,2},
Nagayoshi Ohashi\altaffilmark{3},
Tyler L. Bourke\altaffilmark{4},
Naomi Hirano\altaffilmark{3},
Paul T. P. Ho\altaffilmark{3,4},
Jes K. J{\o}rgensen\altaffilmark{4}, 
Yi-Jehng Kuan\altaffilmark{3,5},
David J. Wilner\altaffilmark{4},
Sherry C. C. Yeh\altaffilmark{3}}

\altaffiltext{1}{Harvard-Smithsonian Center for Astrophysics, 
Submillimeter Array Project, 645 North A'ohoku, Hilo, HI 96720, U.S.A.}
\altaffiltext{2}{Present Address: National Astronomical Observatory of
Japan, ALMA Project Office, Osawa 2-21-1, Mitaka, Tokyo, 181-8588, Japan,
e-mail: s.takakuwa@nao.ac.jp}
\altaffiltext{3}{Academia Sinica Institute of Astronomy and Astrophysics,
P.O. Box 23-141, Taipei 106, Taiwan} 
\altaffiltext{4}{Harvard-Smithsonian Center for Astrophysics, 
60 Garden Street, Cambridge, MA 02138, U.S.A.}
\altaffiltext{5}{Department of Earth Sciences, National Taiwan Normal
University, 88 Sec.4, Ting Chou Rd., Taipei 116, Taiwan}

\begin{abstract}
With the Submillimeter Array (SMA) we have made high angular-resolution
($\sim$ 1\arcsec = 160 AU) observations of the protobinary system IRAS
16293-2422 in the $J = 4 - 3$ lines of HCN and HC$^{15}$N, and in the
continuum at 354.5 GHz.  The HCN (4--3) line was also observed using the
JCMT to supply missing short spacing information.  Submillimeter continuum
emission is detected from the individual binary components of Source A in the
South-East and Source B in the North-West with a separation of
$\sim$5$\arcsec$.  The optically-thin HC$^{15}$N (4--3) emission taken with
the SMA has revealed a compact ($\sim$ 500 AU) flattened structure (P.A. =
-16\degr) associated with Source A.  This compact structure shows a
velocity gradient along the projected minor axis, which can be interpreted
as an infalling gas motion.  Our HCN image including the short-spacing
information shows an extended ($\sim$ 3000 AU) circumbinary envelope as
well as the compact structure associated with Source A, although the P.A.
of the compact structure (= -45\degr) seen in the HCN emission is slightly
different from that of the HC$^{15}$N emission.  A toy model consisting of
a flattened structure with radial infall towards a 1~M$_{\odot}$ central
star reproduces the HCN/HC$^{15}$N position-velocity diagram along the
minor axis of the HC$^{15}$N emission.  In the extended envelope there is
also a North-East (Blue) to South-West (Red) velocity gradient across the
binary alignment, which is likely to reflect gas motion in the swept-up
dense gas associated with the molecular outflow from Source A.  At Source
B, there is only a weak compact structure with much narrower line widths
($\sim$ 2 $\kms$) seen in the optically-thin HC$^{15}$N emission than that
at Source A ($>$ 10 $\kms$), and there is no clearly defined bipolar
molecular outflow associated with Source B.  These results imply the
different evolutionary stages between Source A and B in the common
circumbinary envelope.  Our study demonstrates the importance of adding
short spacing data to interferometer data in order to probe the detailed
structure and kinematics from extended ($>$ 3000 AU) envelopes to inner
compact ($<$ 500 AU) structures around low-mass protostars.
\end{abstract}
\keywords{ISM: individual (IRAS 16293-2422) --- ISM: molecules --- stars:
disk --- stars: formation}

\section{Introduction}

Millimeter molecular-line observations have found 2000--10000 AU scale
molecular envelopes around low-mass protostars
\cite{ben89,zho89,bla94,zho94,bla95,hog97,hog99,tak00,ta03b}, and detailed
interferometric studies in millimeter lines have revealed rotating and
infalling gas motion in these envelopes \cite{oha96,oh97a,oh97b,mom98}.
The dense and warm innermost ($\leq$ 500 AU) region of these low-mass
protostellar envelopes is a likely site of formation of protoplanetary
disks around protostars \cite{ter93,bec96}.  However, there is not a great
deal of information published about the connection between large-scale
protostellar envelopes and small-scale disks around the central protostars.
With the advent of interferometers in the millimeter and centimeter
wavelengths and the use of multiple spectral lines, progress is being made
toward separating the warm and dense regions from the overlying
lower-density and colder gas along the line of sight
\cite{mun90,mom98,ful00,hog01,tak06}.  Recent arcsecond or
subarcsecond-resolution millimeter interferometric observations have
enabled us to investigate the innermost regions of protostellar envelopes
\cite{loo00,bel04,bot04}.  Detailed submillimeter line observations are
also useful for these studies, since submillimeter molecular lines such as
HCN (4--3) trace higher densities ($>$ 10$^{\rm 8}$ cm$^{\rm -3}$) and
temperatures ($>$ 40 K). Earlier studies of low-mass protostellar envelopes
with the Submillimeter Array (SMA)\footnote{The Submillimeter Array (SMA)
is a joint project between the Smithsonian Astrophysical Observatory and
the Academia Sinica Institute of Astronomy and Astrophysics, and is funded
by the Smithsonian Institution and the Academia Sinica.} have demonstrated
that submillimeter interferometric observations are uniquely suited to
probe the innermost part of low-mass protostellar envelopes
\cite{kua04,tak04,cha05,bou05,jor05}.  We note that in order to study the
connection between the large-scale envelopes and inner disks, observations
that sample adequately the emission from large to small scales with high
enough resolution are needed \cite{gue96,wel00,tak04,tak06}.

IRAS 16293-2422 (hereafter I16293) is a Class 0 protobinary system with a
projected separation of $\sim$ 800 AU (Source A and Source B)
\cite{woo89,mun92,loo00} in the Ophiuchus Molecular Cloud, surrounded by an
$\sim$ 8000 AU-scale circumbinary envelope \cite{sch02,sta04}.
Previous SMA observations reported by Chandler et al. (2005) further
demonstrate that one of the binary components, Source A, may be composed of
at least three components within a 1$\arcsec$ region. It has been
reported that there is rotating and infalling gas motion in the
circumbinary envelope of I16293 \cite{wal86,men87,zho95,nar98,ce00a,sch02}.
Large-scale
quadrupolar molecular outflows have been observed in I16293
\cite{wal88,miz90,cas01,hir01,gar02,sta04}. Previous interferometric
\cite{bot04,kua04,sch04,cha05,hua05} and single-dish molecular-line
observations \cite{bla94, dis95, ce00b, sch02, caz03, sch04} have revealed
a number of complex organic molecules and hot-core chemistry in I16293. 
Previous published SMA observations \cite{cha05} have studied in detail the 
structure and kinematics in the circumbinary envelope and circumstellar
disks for each binary companion, the driving sources of the outflows, and
the evolutionary status of each member of the binary.

In this paper, we describe results of HCN (4--3), HC$^{15}$N (4--3), and
354.5 GHz continuum observations of I16293 made with the SMA and James
Clerk Maxwell Telescope (JCMT). This paper focuses on sampling, as
completely as we can, the structure and kinematics from the large-scale
($\sim$ 3000 AU) circumbinary envelope traced by JCMT to the smaller-scale
($\leq$ 500 AU) structures traced with the SMA. For comparison with
previous studies we assume a distance of 160 pc to the Ophiuchus cloud
\cite{whi74}, but we note that more recent studies locate the cloud at a
nearer distance of 120--140 pc \cite{geu89,knu98}.

\section{Observations}
\subsection{Submillimeter Array Observations}

With the Submillimeter Array (SMA) we made HCN ($J$ = 4--3; 354.5055 GHz),
HC$^{15}$N ($J$ = 4--3; 344.200122 GHz), and 354.5 GHz continuum
observations of I16293 on 2003 March 14, July 12, and 2004 June 19. Details
of the SMA are described by Ho, Moran, \& Lo (2004). The SMA is a
double-sideband instrument, and the HCN and HC$^{15}$N lines were observed
simultaneously in different sidebands.  We assigned a spectral window of
the SMA correlator (``chunk'') with a spectral resolution of 0.17 \kms\ to
the HCN line, and a chunk with a resolution of 0.71 \kms\ to the HC$^{15}$N
line\footnote{Due to the instrumental constraint of the SMA correlator, it
is not possible to asign the high-resolution chunk to both the HCN and
HC$^{15}$N lines.}.  Table 1 summarizes the observational parameters.  The
observations were made in three different array configurations to provide
well-sampled ($u,v$) coverage.  The range of the baseline length projected
on the sky was from $\sim$ 10 k$\lambda$ to $\sim$ 223 k$\lambda$, and our
observations were insensitive to structures more extended than $\sim$
16$\farcs$5 ($\sim$ 2600 AU) at the 10\% level \cite{wil94}.  We confirmed
that the visibility amplitudes of the continuum emission from I16293 from
the three observing periods were consistent within the noise level.  The
overall flux uncertainty was estimated to be $\sim$ 30 $\%$.  Part of the
2003 data set has already been published by Kuan et al.\ (2004) and Huang
et al. (2005), which is of lower angular resolution than the full data set
presented here.

The raw visibility data were calibrated and flagged with MIR, which is an
IDL-based reduction package adopted for the SMA from the MMA software
package developed originally for the Owens Valley Radio Observatory
\cite{sco93}. The calibrated visibility data were Fourier-transformed and
CLEANed with MIRIAD to produce images \cite{sau95}.
The spatial resolution is 1$\farcs$1 $\times$ 0$\farcs$6 (P.A. = 39\degr),
1$\farcs$3 $\times$ 1$\farcs$2 (P.A. = 30\degr), and
1$\farcs$6 $\times$ 1$\farcs$3 (P.A. = 14\degr),
in the 354.5 GHz continuum, HCN, and HC$^{15}$N images, respectively.
The spatial resolution of the HCN image is set to be the same as that of
the combined SMA + JCMT image (see next section), in order to
directly compare the SMA and
combined images and to see the effect of missing short-spacings.

\placetable{tlb-1}

\subsection{James Clerk Maxwell Telescope Observations}

Submillimeter single-dish mapping observations of I16293 in the HCN (4--3)
line were made with the James Clerk Maxwell Telescope (JCMT) on 2004 March
7 and April 7. The JCMT beam size at 354 GHz band was $\sim$ 15$\arcsec$
and the typical system temperature was 600 K. We observed a 7 $\times$ 7
map centered on the field center of the SMA observations at a grid spacing
of 7$\farcs$5, providing a Nyquist-sampled map in the 45$\arcsec$ $\times$
45$\arcsec$ region. Each position in the map was observed for 1 minute and
the map was repeated 6 times. The central map position was observed at the
start, middle and end of each map as a check on the relative flux
calibration. The observations were made in dual-polarization mode, and the
2 polarizations averaged, resulting in an rms noise level per 0.13 km
s$^{-1}$ channel of $\sim$ 0.13 K in T$_{A}^{*}$. Pointing and focusing
were checked before each set of maps.  The conversion factor from
T$_{A}^{*}$ (K) to $S$ (Jy beam$^{-1}$) was derived to be 36.7 as;

\begin{equation} 
S = \frac{2k_{B}\Omega_{beam}}{\lambda^2}\frac{T_{A}^{*}}{\eta_{mb}},
\end{equation} 

where $k_{B}$ is the Boltzmann constant, $\Omega_{beam}$ is the solid angle
of the JCMT beam (= 15$\arcsec$), $\lambda$ is the wavelength, and
$\eta_{mb}$ is the main beam efficiency of the JCMT (= 0.63).  The mapped
region covers well the field of view of the SMA observations ($\sim$
35$\arcsec$), which enables us to compare the JCMT flux to the SMA flux.
The SMA observations recovered $\sim$ 35$\%$ (Source A) and 31$\%$ (Source
B) of the total HCN(4-3) flux observed with the JCMT toward the binary
system (The JCMT observations do not resolve the 5\arcsec\ binary).  We
combined the JCMT data with the SMA data, and in the subsequent sections we
will mainly discuss the combined images. The details of the combining
process are described in Appendix.  The resultant synthesized beam size and
the rms noise level in the combined images is 1$\farcs$3 $\times$
1$\farcs$2 (P.A. = 30\degr) and $\sim$ 0.90 Jy beam$^{-1}$ channel$^{-1}$,
respectively, where the velocity resolution is the same as that of the SMA
data.

\section{Results}
\subsection{Submillimeter Continuum Emission}

In Figure 1 we show our 354.5 GHz continuum image of I16293.  Previous SMA
images of the continuum emission in I16293 at somewhat lower angular
resolutions have been presented in Kuan et al. (2004) and Chandler et al.
(2005; although they subsequently improved their resolution through
``super-resolution'' imaging techniques, discussed below).  Submillimeter
continuum emission from the individual binary protostars, called Source A
(south-east) and B (north-west) \cite{mun92}, are evident.
The total continuum fluxes at 354.5 GHz are 3840 and 4050 mJy at Source A
and B,
respectively, which are slightly higher than the continuum fluxes at 305
GHz (3460 and 3150 mJy at Source A and B; Chandler et al. 2005).  Chandler
et al. (2005) reported that the spectral indexes ($\equiv$ $\alpha$) in the
millimeter and submillimeter regime are 2.91 and 2.51 at Source A and B,
respectively.  We re-estimated the spectral indexes including our new
submillimeter measurements, and the values are 2.87 and 2.68 at Source A
and Source B, respectively.  These values are consistent with the values by
Chandler et al. (2005) within the uncertainties ($\sim$ $\pm$ 0.1).
The submillimeter continuum emission at Source A shows evidence of
extension along the North-East to South-West direction, while that at
Source B is more compact.  In Table 2, we summarize the properties of the
submillimeter continuum emission from both protostars after the
deconvolution of the synthesized beam.  Using ``super-resolution'' imaging
by only using data with $uv$ distance greater than 55 $k\lambda$, the
extension at Source A is resolved into two components with the SMA (Aa and
Ab; Chandler et al.\ 2005), possibly a close binary system.  The peak
positions of the 354.5 GHz continuum emission are slightly ($\sim$
0$\farcs$4) offset from the 115 GHz \cite{loo00} and 300 GHz \cite{cha05}
continuum positions, presumably due to the limited calibration accuracy of
our higher-frequency observations.  Detailed multi-wavelength (230-690 GHz)
analysis of the submillimeter continuum emission will be presented in a
forthcoming paper.

\placefigure{f1}
\placetable{tlb-2}

\subsection{Spatial and Velocity Distribution of the HCN and HC$^{15}$N
Emission}

In Figure 2, we compare the HCN (4--3) line profiles from SMA and combined
SMA+JCMT data at Source A and B, along with the SMA HC$^{15}$N (4--3) line
profiles.  We also show the average (Source A + Source B) SMA and SMA+JCMT
spectra for comparison with the JCMT HCN spectrum.  The HCN (4--3) line
profiles show two emission peaks with a brighter blue-shifted peak and an
absorption dip.  The SMA and combined HCN spectra show a larger brightness
temperature with more prominent wing-like emission as compared to those
obtained with the JCMT, while the interferometric-only spectra
systematically miss lower-velocity emission around $V_{LSR}$ = 2 - 4 km
s$^{-1}$ as compared to the combined spectra, particularly at Source A.
The dip in the SMA+JCMT and the SMA spectra seems to show absorptions
against the continuum.  The HC$^{15}$N spectra shown in Figure 2 include
new data taken on June 19, 2004 as compared to the HC$^{15}$N spectra shown
by Kuan et al. (2004), and the spatial resolution ($\sim$ 1$\farcs$6
$\times$ 1$\farcs$3) is higher than that of Kuan et al. (2004) ($\sim$
2$\farcs$7 $\times$ 1$\farcs$3).  At Source A, the HC$^{15}$N line profile
shows a single peak near the dip velocity.  The centroid velocity of the
HC$^{15}$N line at Source A is estimated to be $\sim$ 2.9 km s$^{-1}$ from
a single-component Gaussian fitting to the spectrum.  At Source B, there is
a weak ($\sim$ 4$\sigma$) HC$^{15}$N emission near the centroid velocity.
This velocity is slightly bluer than the HCN dip velocity ($\sim$ 4.3 km
s$^{-1}$), although the coarser velocity resolution of the HC$^{15}$N data
($\sim$ 0.71 km s$^{-1}$) prevents us from making the direct comparison
between the HCN and HC$^{15}$N velocities.  Taking into account the coarse
velocity resolution in the HC$^{15}$N line, we adopt the symmetric velocity
of the HCN Position-Velocity diagram (Figure 7) as a systemic velocity
($\sim$ 3.6 km s$^{-1}$).  At this velocity, the HCN (SMA+JCMT)/HC$^{15}$N
(SMA) line intensity ratio toward Source A is $\sim$ 4, which implies an
optical depth of $\sim0.3$ for the HC$^{15}$N line assuming an
N$^{14}$/N$^{15}$ isotopic ratio of 270 \cite{luc98}.  Therefore, the
HC$^{15}$N line is likely to be optically thin.  The redshifted
self-absorption in the HCN spectra against the continuum at Source A could
be a sign of the inward motion toward the continuum source
\cite{zho92,mye95,nar98,dif01,sch02,sta04}, although the presence of the
extended HCN absorption in the entire region (See Figure 4) implies
significant contamination from extended low-temperature material.  The
red-shifted self-absorption from the systemic velocity can be reproduced
using models of infalling cores with the Larson-Penston flow
\cite{lar69,pen69}, as has been shown by Masunaga \& Inutsuka (2000). Such
a red-shifted dip was also detected in B335 \cite{cho99,eva05}, which is
one of the most well-studied infalling cores.

Figure 3 shows two different HCN integrated intensity maps; one made using
only the SMA data (left), and the other made using the JCMT and SMA
combined data (middle). Figure 3 also shows the HC$^{15}$N integrated
intensity map taken with the SMA for comparison (right).  The continuum
emission has been subtracted before forming these maps.  In the
optically-thin HC$^{15}$N emission there is a compact emission associated
with Source A. From a 2-dimensional Gaussian fitting to the image, the FWHM
size of this compact HC$^{15}$N emission after the deconvolution of the
synthesized beam is 450 AU $\times$ 250 AU (P.A. = -16\degr).  A similar
compact structure is also seen in the SMA and combined SMA+JCMT HCN images
with a size of 630 AU $\times$ 320 AU (P.A. = -43\degr) and 740 AU $\times$
440 AU (P.A. = -45\degr), respectively, although the P.A. of the HCN
component is slightly different from that of the HC$^{15}$N emission.  The
HCN emission also shows extensions to the North-West and to the North-East
of Source A in addition to this compact structure.  Importantly, the
combined HCN image shows that all of these features are surrounded by a
halo component with a size of $\sim$ 3000~AU, demonstrating that the SMA is
not sensitive to the extended structure. This halo structure is most likely
a molecular envelope surrounding the I16293 protobinary system.  Similar
extended structures are also seen in the H$_{2}$CO (4$_{1,3}$--3$_{1,2}$)
and SO (7$_{7}$--6$_{6}$) emission, although the detailed distributions are
somewhat different from each other \cite{cha05}.

In Figure 4, we present velocity channel maps of the SMA+JCMT HCN data.  At
high blueshifted velocities ($-$2.8 \kms\ $< V_{LSR} <$ -1.2 \kms), there
is a compact gas component associated with Source A. This high-velocity
component corresponds to the compact structure seen in the total integrated
intensity map of Figure 3.  From $V_{LSR}$ = -0.73 \kms, a secondary
component is seen at the south-east of Source B.
From $V_{LSR}$ $\sim$ 1.9 \kms, an extended halo component appears until
$V_{LSR}$ = 4.4 \kms, where little emission is evident
due to the extended self-absorption in the HCN emission.
After this velocity the redshifted halo component appears again.
From $V_{LSR}$ $\sim$ 6.5 \kms a compact gas component toward Source A is
seen, which corresponds to the redshifted counterpart of the compact
structure at Source A.

In order for us to see the systematic velocity structures more clearly, we
integrated these velocity channel maps into four different velocity
regimes, that is, low-velocity blueshifted (2.5 -- 3.4 km s$^{-1}$) and
redshifted (5.0 -- 6.5 km s$^{-1}$), and high-velocity blueshifted (-2.6 --
0.7 km s$^{-1}$) and redshifted (7.4 -- 9.1 km s$^{-1}$) emission.  In
Figure 5, we compare these four HCN velocity channel maps to the CO (2--1)
maps taken with the SMA \cite{she06}, integrated into the same four
velocity regimes. The spatial resolution in the CO (2--1) map is 3$\farcs$3
$\times$ 2$\farcs$0 (P.A. = 40\degr).  In the low-velocity regimes, the
overall distributions of the HCN emission, which shows extended halo
structures, are very similar to those of the CO (2--1) emission.  This
suggests that the extended halo structures are affected by a low-velocity
outflow or turbulent gas traced by CO.  The redshifted halo component
appears mostly at the south-west and south-east of the binary, while the
blueshifted halo component at the north-east and north-west of the binary.
In the high velocity regimes, the HCN emission shows extensions as well as
the compact structures associated with Source A.  The comparison with the
CO maps shows that the HCN extensions have distributions similar to those
of CO, particularly at the high redshifted velocity. This suggests that the
HCN extensions are also affected by the high-velocity outflow.  On the
other hand, the compact HCN emission associated with Source A seems to show
different distributions from those of CO, which suggests that it has
different origin. The compact HCN component is also
distinct from the more extended low-velocity emission.  We will discuss the
origin of these different velocity components in the next section.

\placefigure{f2}
\placefigure{f3}
\placefigure{f4}
\placefigure{f5}

\section{Discussion}
\subsection{Origin of the Different Velocity Structures}

Our SMA and JCMT observations have revealed a high-velocity compact ($\sim$
500 AU) structure associated with Source A and an extended ($\sim$ 3000 AU)
low-velocity circumbinary envelope around sources A and B.  In this
subsection, the origin of these different velocity structures is discussed.

In Figure 6, we compare the high-velocity blueshifted and redshifted HCN
and HC$^{15}$N emission to the CO (2--1) outflow map taken with the SMA
(Yeh et al. 2007).  As discussed in Yeh et al. (2007), source A is the most
likely to be driving the E-W CO outflow, while the driving source of the
NW-SE outflow remains uncertain.  In the optically-thin HC$^{15}$N
emission, the high-velocity blueshifted and redshifted components exhibit a
velocity gradient along the minor axis (P.A. 74\degr) of the compact
HC$^{15}$N structure.  The velocity gradient along the minor axis is
approximately parallel to the E-W CO outflow,
though the position angle of the E-W CO outflow can not be
well-defined due to the wide opening angle.  Although the direction
of the minor axis of the
compact HCN emission is slightly different from that of the compact
HC$^{15}$N emission, the direction of the velocity gradient in the HCN
emission is similar to that of the HC$^{15}$N emission.
One possible interpretation of the observed velocity gradient in the
compact flattened structure is outflowing gas, because both the CO
outflow and the HCN/HC$^{15}$N emission show the same trend of velocity gradient.
In fact, the extended HCN emission is strongly affected by the molecular
outflow, as has been discussed in the previous section.  The slight
difference of the position angle between the compact HCN and HC$^{15}$N
component is also likely to be due to the contamination from the extended
component.
However, the optically-thin HC$^{15}$N emission, which should
be less affected by the extended outflow component,
is clearly elongated in a different direction from the outflow direction (Figure 5 and 6),
suggesting that the velocity gradient is unlikely to be due to the outflow.
We interpret the
observed velocity gradients in the HCN and HC$^{15}$N emission to be an
infalling gas motion in the flattened disklike structure toward Source A
\cite{hay93,oha96,mom98}.

In the left panel of Figure 7, we show position-velocity (P-V) diagrams
along the minor axis of the HC$^{15}$N compact emission at Source A (see
Figure 6), in the HCN emission with the SMA-only (gray) and SMA+JCMT (black
contour), and in the HC$^{15}$N emission with the SMA (red contour).  In
the P-V diagrams, both the HCN and HC$^{15}$N lines trace the
higher-velocity emission located close to Source A, and this velocity
structure corresponds to the compact emission at Source A.  We made a
number of model P-V diagrams of a geometrically thin disk with Gaussian
intensity distribution to reproduce this component, one of which is shown
in the right panel of Figure 7. From the estimated major and minor axis in
the SMA HC$^{15}$N image we estimated the inclination angle of the
flattened structure from the plane of the sky to be $\sim$ 57\degr
($\equiv$ cos$^{-1}$(maj/min)), and we adopted this inclination in the
model P-V diagram. We also assumed the internal velocity dispersion
($\equiv$ $\sigma$) to be 1.0 km s$^{-1}$ \cite{sta04}. On these
assumptions, we made model P-V diagrams with different central masses, and
the model P-V diagram in Figure 7 shows that the compact high-velocity
emission could be interpreted as an infalling disk onto the central stellar
mass of $\sim$ 1 M$_{\odot}$ with an acceptable mass range from $\sim$ 0.5
M$_{\odot}$ to $\sim$ 2.0 M$_{\odot}$.  An uncertainty of $\pm$10\degr of
the inclination angle of the flattened structure could reproduce additional
$\sim$ 20$\%$ uncertainty of the central mass. From VLBI studies of
H$_{2}$O masers in I16293, Imai et al. (1999) have found a
rotating-infalling disk with an outer radius of 100 AU around Source A.
Their estimates of the inclination angle and the central stellar mass are
$\sim$ 35\degr and $\sim$ 0.3 M$_{\odot}$, respectively. These estimates
are roughly consistent with our estimates ($\sim$ 57\degr, and $\sim$ 1.0
M$_{\odot}$), and the compact HCN structure detected with our observations
can be interpreted as tracing the outer part of the same infalling disk.

From the optically-thin HC$^{15}$N emission, we estimated the mass of the
compact structure within the radius of $\sim$ 230 AU ($\equiv$ $r$) to be
0.084 M$_{\odot}$ ($\equiv$ $M_{r}$). Here, we assume the HC$^{15}$N
emission to be optically thin, the excitation temperature to be 30 (K)
\cite{sch02}, and the HC$^{15}$N abundance to be 7.4 $\times$ 10$^{-11}$
\cite{kua04}.  The infalling velocity at that radius is 2.8 km s$^{-1}$
($\equiv$ $v_{r}$) from our model. Then, the mass accretion rate ($\equiv$
$\dot M$) can be estimated from 

\begin{equation} 
\dot M = M_{r}v_{r} / r
\end{equation} 

and we find a value of 2.2 $\times$ 10$^{-4}$ (M$_{\odot}$ year$^{-1}$).
This value of the mass accretion rate is 4 times higher than that estimated
by Sch{\"o}ier et al. (2002) ($\sim$ 5 $\times$ 10$^{-5}$ M$_{\odot}$
year$^{-1}$). Schoier et al 2002 assumed that the entire ($\sim$ 3000 AU in
size) circumbinary envelope is infalling and derived the mass accretion
rate from a model fitting to their JCMT spectra.  On the other hand, from our
higher spatial-resolution observations we suggest that only the compact
HCN/HC$^{15}$N component associated with Source A is infalling, and we
derived the mass accretion rate in the compact HCN component from our toy
model. The difference of the estimated mass accretion rate probably arises
from the different configuration adopted.  From the estimated mass
accretion rate, the accretion luminosity ($\equiv$ $L_{acc}$) can be
calculated as 

\begin{equation} 
L_{acc} = \frac{GM_{*}\dot M}{R_{*}}
\end{equation} 

where $G$ is a gravitational constant, $M_{*}$ is a mass of the central
protostar, and $R_{*}$ is a radius of the central protostar. If we assume
$R_{*}$ = 4 $R_{\odot}$ \cite{sta80}, $L_{acc}$ is estimated to be $\sim$
1600 $L_{\odot}$ with our $\dot M$ value and to be $\sim$ 390 $L_{\odot}$
with the $\dot M$ value by Sch{\"o}ier et al.  (2002).  These values are
much higher than the bolometric luminosity of I16293 (= 27 $L_{\odot}$;
Walker et al. 1986), so that the so-called ``luminosity problem'' (Kenyon
et~al.\ 1990) discussed by Ohashi et al.\ (1996) and Saito et al.\ (1996)
for L1551 IRS5 also exists in I16293.  The reason of this discrepancy is
still unclear. One possible explanation is that the infalling motion found
in the present study is not onto the surface of the protostar itself, but
onto the central rotationally-supported disk around the protostar.  Here,
the radius of the terminal point in the gravitational potential is likely
to be much larger than that adopted in the above estimates. If we assume
the radius of the terminal point to be $\sim$ 1 AU, our estimated mass
infalling rate is consistent with the bolometric luminosity. Another
possible explanation is that the accretion may be non-steady.

We note that an infalling motion has the same radial dependence as that of
a Keplerian rotation ($\propto$ $r^{-0.5}$), and that they are
indistinguishable from the examination of the P-V diagram alone. The only
clue to distinguish the two motions is the disk orientation as compared to the
direction of the velocity gradient.  In fact, Huang et al. (2005) have
interpreted the velocity gradient in the HC$^{15}$N emission as a Keplerian
rotation around Source A.  However, the position angle of the flattened
structure in the optically-thin line (-16\degr) as well as that of the
inner rotating-infalling disk traced by the H$_{2}$O maser \cite{ima99},
seems to favor the interpretation of the infalling motion. Furthermore, the
interpretation of the Keplerian rotation implies that there is a
well-developed 500 AU scale rotationally-supported disk around the Class 0
protostar with the significant molecular outflows. In general, such
well-developed large-scale ($\sim$ 500 AU) Keplerian disks are observed
around more evolved sources, that is, Class II sources, as in the case of a
circumbinary Keplerian ring around GG Tau \cite{gui99} and a Keplerian
protoplanetary disk around DM Tau \cite{gui98}.  From these considerations,
we suggest that the infall gas motion is a more appropriate interpretation
than that of the Keplerian rotation.

The SMA+JCMT P-V diagram in Figure 7 exhibits another lower-velocity,
extended component as seen in the velocity channel maps of Figure 4, which
is mostly filtered out by the interferometer alone. The velocity structure
in this extended lower-velocity component is clearly different from that in
the compact high-velocity component; the velocity increases further from
Source A.  Since the cut of the P-V diagram is approximately parallel to the axis of the
associated CO outflow (see Figure 6) and this type of velocity gradients is
expected in molecular outflows driven by the wide-angle wind, we suggest
that this extended HCN emission arises from turbulent gas swept up by the
associated outflow from Source A. Our simple thin-layer model with the
Gaussian intensity distribution shown in the right panel of Figure 7 is
consistent with the interpretation that this low-velocity component could
be an outflowing gas perpendicular to the compact high-velocity structure.

In Figure 8, we show P-V diagrams along the cut through Source A and B
(P.A. = 138.8\degr) in the HCN emission with the SMA-only (gray) and
SMA+JCMT (black contour), and in the HC$^{15}$N emission with the SMA (red
contour).  At the position of Source A the wide velocity width in the HCN
and HC$^{15}$N high-velocity component is evident.  At the position of
Source B, a weak HC$^{15}$N component with a narrow line width ($\sim$ 2
\kms) is seen, which is smeared out in the total integrated intensity map
of Figure 3 and embedded in the extended HCN emission.  This component at
Source B is most likely a molecular structure found by the previous SMA
observations \cite{kua04,cha05}, as well as that by the previous Plateau de
Bure observations \cite{bot04}. In Figure 8, there is also an extended
structure with narrow line width, which is likely to be the extended
ambient gas.

\placefigure{f6}
\placefigure{f7}
\placefigure{f8}

\subsection{Different Evolution of the Binary Protostars ?}

Our submillimeter HCN and HC$^{15}$N observations of I16293 show that an
intense compact ($\sim$ 500 AU) molecular component with a wide velocity
width ($>$ 10 \kms) is associated with Source A but only a weak molecular
component with a narrow velocity width ($\sim$ 2 \kms) is associated with
Source B.
SMA observations of I16293 in the CO (2--1, 3--2) lines have revealed that
Source A is clearly driving molecular outflows, and that there is no clear
outflow activity from Source B \cite{she06}.  Source A is associated
with a number of hot core tracers such as complex organic molecules that
are less prominent or not present at Source B \cite{kua04,cha05}.
These results imply that Source A and B are at different evolutionary
stages in the common circumbinary envelope, although it is unclear which
source is younger.  In terms of protostellar activity traced by molecular
lines, Source A seems to be younger than Source B, since Source A has
substantial surrounding molecular gas and clear outflow emission, and shows
a number of hot core tracers \cite{kua04,cha05}, while Source B is more
like a T Tauri star without a significant gas component or CO outflow. On
the other hand, if the compactness in the continuum emission represents
younger evolutionary stage of the circumstellar disk, Source B may be
younger than Source A as suggested by Rodr\'{\i}guez et al. (2005) and
Chandler et al (2005).  In fact, Chandler et al (2005) have found
redshifted SO (7$_{7}$--6$_{6}$) absorption against the strong continuum
emission at Source B, which they suggest is the unambiguous detection of
infall toward the central protostar.

There are other examples of binary protostars where the components exhibit
different observational characteristics that suggests they are in different
evolutionary stages.  Bourke (2001) has found two 7 $\mu$m sources with a
separation of $\sim$ 3400 AU in BHR 71 with the Infrared Space Observatory.
The brighter companion (IRS~1) is associated with the millimeter continuum
source (BHR 71 mm) while the other weaker companion (IRS~2) is not, and IRS
1 drives a powerful molecular outflow while IRS 2 drives a much less
massive outflow. From these results, it is suggested that IRS1 and IRS2 in
BHR 71 is a protobinary system at different evolutionary stages, and that
IRS 2 is more evolved than IRS 1. High angular-resolution ($\sim$
2$\arcsec$) observations of CB 230 show compact millimeter emission
associated with only one component of an NIR protostellar pair separated by
$\sim$ 10$\arcsec$, suggesting that like BHR 71 only this component has a
substantial circumstellar disk, although both protostars in CB 230 drive CO
outflows \cite{lau01}. A similar example is also seen in the SVS 13 close
binary system (separation $\sim$ 65 AU). Anglada et al. (2004) reported
from their high-resolution 7-mm observations with VLA that only one of the
components of the SVS 13 system (VLA 4B) is associated with detectable
circumstellar dust emission, while the other component is optically visible
without significant dust emission. These recent observational results
indicate that members of protostellar binary systems can exist at different
evolutionary stages.

Theoretically it is not clear how to form binary companions at different
evolutionary stages in the common envelope and this idea remains
controversial.
According to recent theoretical models of binary formation in this common
envelope type \cite{nak03}, fragmentation of ``the first bar'' which was
made through the contraction of the envelope produces binary or multiple
stars, but the stellar age must be the same. Subsequent merging of the
fragments may be necessary to make binary companions at different
evolutionary stages. Ochi et al. (2005) have conducted high-resolution
numerical simulations of an accretion from a circumbinary envelope onto the
primary and secondary of the binary companion. Their simulations suggest
that the accretion rate of the primary is larger than that of the
secondary, regardless of the specific angular momentum of the accreting
gas, and that the gas accretion tends to increase the mass difference
between the primary and secondary. Their results are qualitatively
different from earlier works by Bate \& Bonnell (1997), which suggests that
the primary accretes more than the secondary only when the accreting gas
has a low specific angular momentum. More theoretical works, as well as an
accumulation of observational data, are required to address this issue.

\section{Conclusions}

We have carried out submillimeter interferometric and single-dish
observations of the well-known protobinary system IRAS 16293-2422 with the
SMA in the HCN (4--3) and HC$^{15}$N (4--3) lines, in the continuum at
354.5 GHz and with the JCMT in the HCN (4--3) line.  These data have
provided us with the following results;

1. The 354.5 GHz continuum observations with the SMA resolve the
individual binary companions (Source A and Source B).  The dust emission at
Source A is elongated along the North-East to South-West direction (300
$\times$ 150 AU; P.A. = 33\degr), while that at Source B is more compact
(150 $\times$ 140 AU).  The elongation at Source A is likely to reflect the
presence of the close binary system (Aa and Ab).  The integrated
submillimeter flux in both sources is $\sim$ 4.0 Jy.

2. The optically-thin HC$^{15}$N (4--3) emission 
reveals a compact ($\sim$ 500 AU) flattened structure (P.A. = -16\degr)
associated with Source A.  There is a clear velocity gradient in this
compact structure along the East (Blue) to West (Red) direction, which can
be interpreted as an infalling gas motion in the flattened structure onto
Source A with a central mass of $\sim$ 1 M$_{\odot}$ and an inclination
angle of $\sim$ 57\degr\ from the plane of the sky.  The combined HCN image
shows an extended ($\sim$ 3000 AU) circumbinary envelope as well as the
compact structure associated with Source A.  In the extended circumbinary
envelope there is also a low-velocity North-East (Blue) to South-West (Red)
gradient, which may be due to gas swept up by the outflow.

3. Contrary to Source A, at Source B there is only a weak molecular
component with a much narrower line width ($\sim$ 2 \kms), and no clear
outflow activity.  These results, together with the fewer number of complex
molecules associated with Source B \cite{kua04,cha05}, may suggest
different evolutionary stages between Source A and B in the common
circumbinary envelope.  With the data presented we are unable to determine
whether Source B is younger or older.

\acknowledgments

We are grateful to R. Kawabe, E. F. van Dishoeck, and P. C. Myers for
fruitful discussions. We also acknowledge T. Hanawa and F. Nakamura for
their comments on theoretical studies of binary formations. We would like
to thank all the SMA staff supporting this work.  S.T. and T.B. were
supported by a postdoctoral fellowship of the Smithsonian Astrophysical
Observatory. The research of Y.-J. K. was supported by NSC
94-2112-M-003-010.



\appendix
\section{Combining SMA and JCMT Data}

The procedure adopted to combine the SMA and JCMT data is similar to that
described by Takakuwa et al. (2003a), which is based on the description for
combining single-dish and interferometric data by Vogel et al. (1984), and on
the MIRIAD scripts developed by Wilner \& Welch (1994).  We first
re-sampled the JCMT image cube along the velocity axis to match the velocity
channels with that of the SMA visibility. Next, we deconvolved the JCMT
image by the 15$\arcsec$ Gaussian image which represents the JCMT beam.
Then, we multiplied the JCMT image with the 35$\arcsec$ Gaussian image
which is intended to approximate the SMA primary beam, that
is, inverse of the primary beam correction.  In these beam corrections we
ignore any effect of sidelobes of the JCMT and SMA beam, which are not
well-known. From the resultant JCMT image cube we made JCMT visibility data
which fill in the central hole of the SMA $uv$-track by the MIRIAD task
$uvrandom$ and $uvmodel$. The single-dish sampling in the $uv$ plane was
chosen so that the number of the $uv$ data points per unit $uv$ distance is
similar to that of the SMA data (= 40 $uv$ data points per 1$k\lambda$).
We checked that the JCMT amplitude is approximately consistent with the SMA
amplitude.

The JCMT visibility data were Fourier-transformed along with the SMA
visibility data to make a combined image cube by the MIRIAD task $invert$.
We adopted robust = 0.5 weighting for the imaging, which seems to provide a
good compromise between the spatial resolution and the noise level.  The
resultant synthesized beam size and the rms noise level is 1$\farcs$3
$\times$ 1$\farcs$2 (P.A. = 30\degr) and $\sim$ 0.90 Jy beam$^{-1}$
channel$^{-1}$, respectively. Negative sidelobes in the SMA synthesized
beam are significantly suppressed in the combined beam.  The final combined
image with the JCMT and SMA has a high spatial dynamic range from
large-scale ($\sim$ 3000 AU) to small scale ($\sim$ 200 AU) without any
effect of missing flux. In Figure 9, we compare the JCMT, SMA, and combined
images of I16293 in the HCN emission.  The JCMT image shows a 3000 AU-scale
circular blob and western elongation centered at Source A without any
fine-scale structure. In the SMA image, finer structures are picked up but
extended gas components are mostly resolved out. In the combined image,
both extended and compact structures are successfully sampled.

\placefigure{f9}

\section{Importance of Combining Single-Dish and Interferometric Data}

Our combined SMA+JCMT observations of I16293 in the submillimeter HCN
emission have revealed that the submillimeter emission extends more than
$\sim$ 3000 AU in I16293.  This extent cannot be recovered with the SMA,
since the minimum projected $uv$ distance of the SMA observations was
$\sim$ 10.0 $k\lambda$ and the SMA observations were insensitive to
structures more extended than $\sim$ 2600 AU at 10$\%$ level \cite{wil94}.
In fact, our SMA observations recovered only $\sim$ 35$\%$ of the total
JCMT flux toward Source A.  For comparison, SMA observations of L1551 IRS5,
another low-mass protostellar envelope, recovered only $\sim$ 11$\%$ of the
total submillimeter CS (7--6) flux, which suggests that there is $>$ 1500
AU-scale extended submillimeter emission in L1551 IRS5 \cite{tak04}.

The rotational energy level of the upper state is 43 K and 66 K in HCN
(4--3) and CS (7--6), and the critical density $\geq$ 10$^{\rm 8}$ cm$^{\rm
-3}$ and $\geq$ 10$^{\rm 7}$ cm$^{\rm -3}$, respectively. These
temperatures and gas densities are much higher than those traced by
millimeter-wave tracers, such as C$^{\rm 18}$O ($J$=1--0) (5 K and
$\sim10^{4}$ cm$^{\rm -3}$; Sargent et al. 1988, Momose et al. 1998) or
H$^{\rm 13}$CO$^{\rm +}$ ($J$=1--0) (4 K and $\sim10^{5}$ cm$^{-3}$; Saito
et al. 2001). It is puzzling why those submillimeter molecular lines which
trace such warmer and denser gas can be so extended. It seems to be
difficult to make gas temperature high enough only by the heating from
central stars (e.g., Lay et~al.\ 1994).  As discussed in $\S$4.1, there is
significant contamination from the associated outflows in the submillimeter
HCN emission, and the interaction between the circumbinary envelope and the
outflowing gas could be a source of such extended submillimeter emission
\cite{ave96,nak00}. To sample those extended submillimeter emission lines
and to study the structure and kinematics from protostellar envelopes ($>$
1500 AU) to compact circumstellar disks ($\leq$ 200 AU) unambiguously, it
is quite important to make both submillimeter single-dish and
interferometric observations and to combine the two data sets.

\clearpage
\begin{figure}
\epsscale{.80}
\plotone{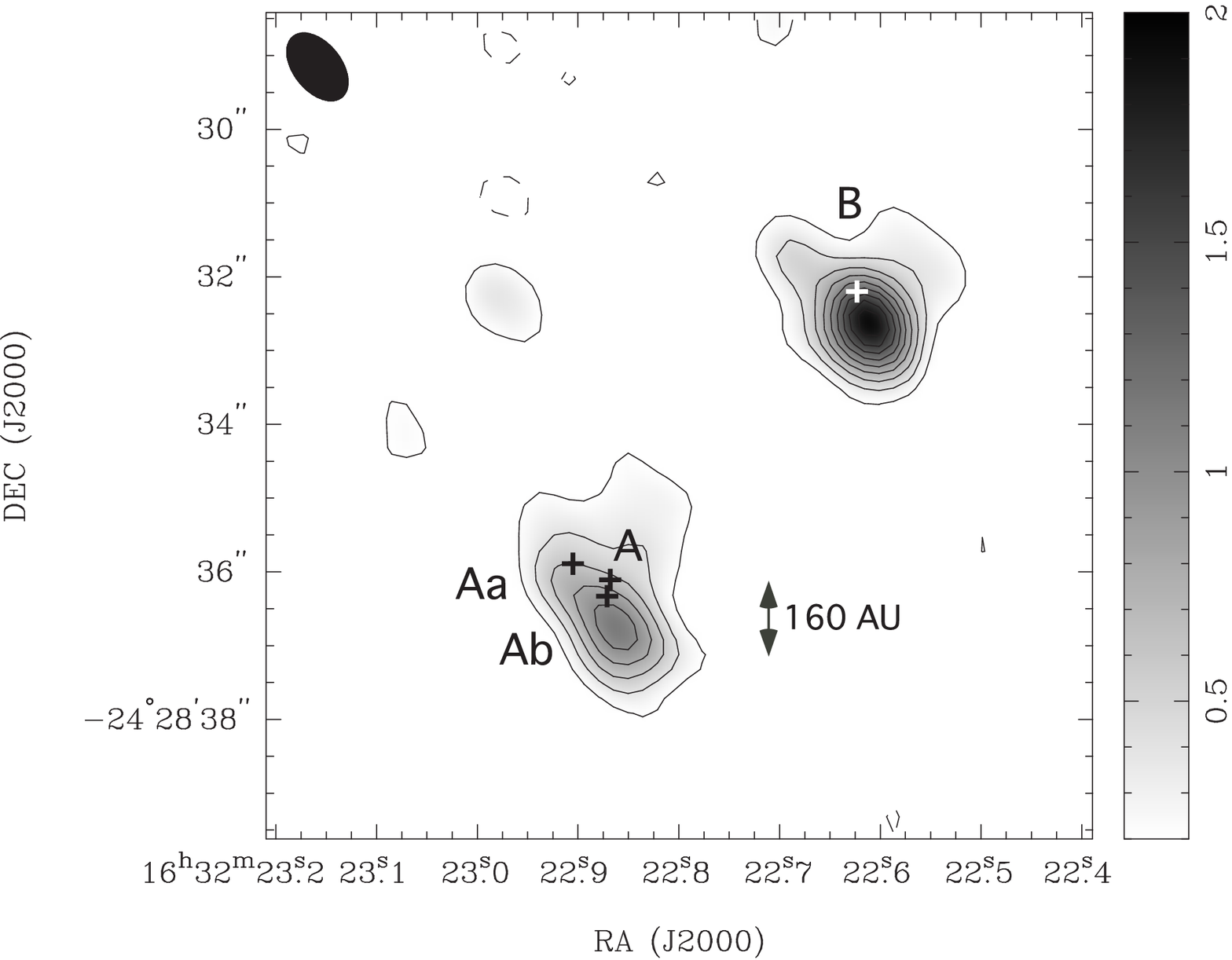}
\caption{354.5 GHz continuum image in I16293 taken with the SMA. Contour
levels are from 3$\sigma$ in steps of 3$\sigma$ (1$\sigma$ = 0.066 Jy
beam$^{-1}$).  The synthesized beam size, shown at the top left corner of
the panel, is 1$\farcs$05 $\times$ 0$\farcs$64 (P.A. = 38.6\degr) with the
uniform weighting. Crosses with the label ``A'' and ``B'' indicate
positions of the 2.7-mm continuum peak observed with BIMA (Looney, Mundy,
\& Welch 2000), and those with the label ``Aa'' and ``Ab'' positions of the
300 GHz continuum peaks observed by Chandler et al. (2005).  The size of
the crosses shows the astrometric accuracy of the BIMA data.  
\label{fig1}}
\end{figure}

\clearpage
\begin{figure}
\includegraphics[angle=90,scale=1.1]{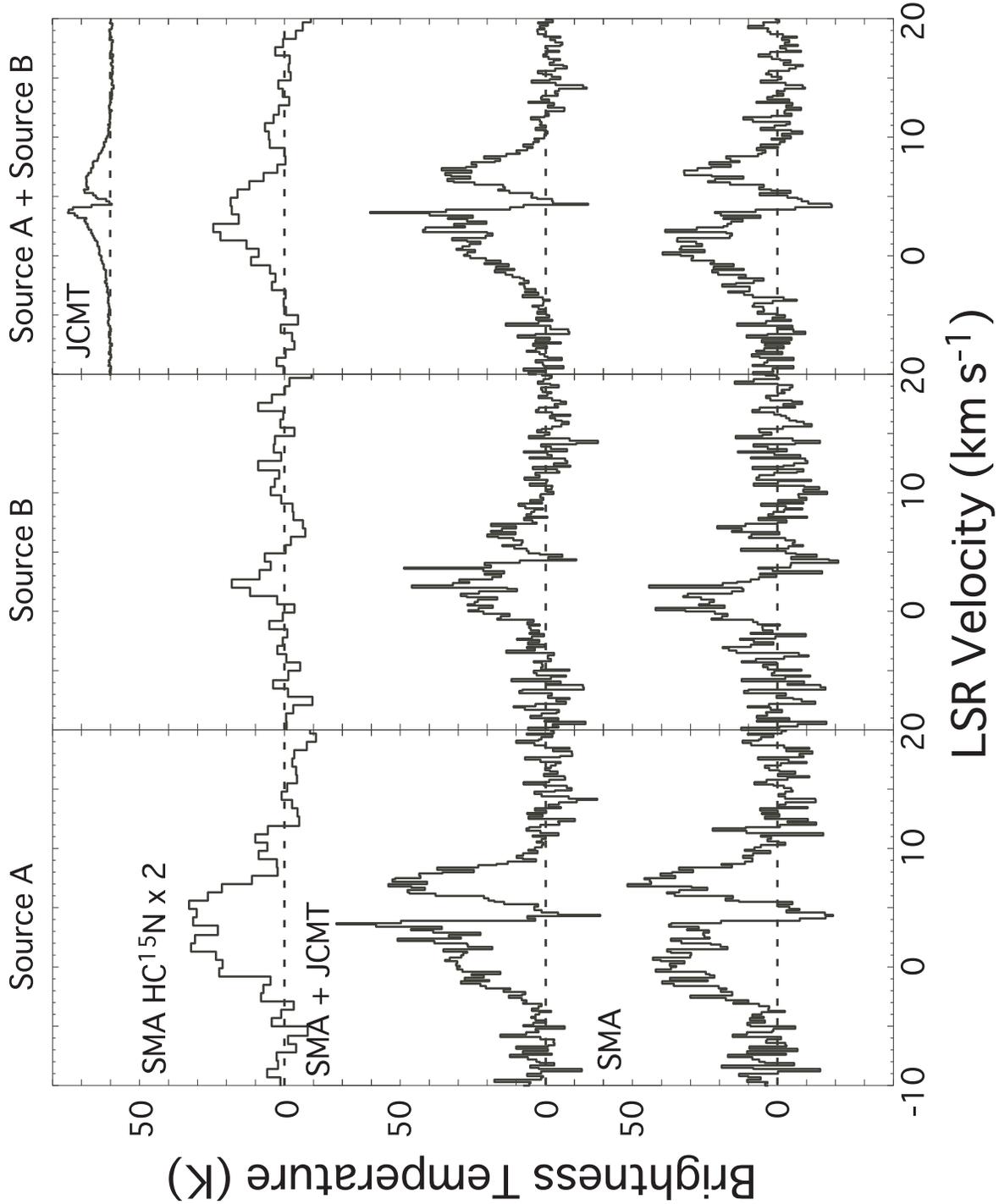}
\caption{HCN (4--3) line profiles toward Source A (left) and Source B
(middle) taken with the SMA (bottom) and SMA + JCMT (middle), and
HC$^{15}$N (4--3) line profiles taken with the SMA (top). For comparison
with the JCMT HCN (4--3) spectrum, we also plot the average spectra toward
Source A and Source B (right). The spatial resolution of the HCN spectra is
1$\farcs$3 $\times$ 1$\farcs$2 (P.A. = 30\degr).
The spatial resolution of the JCMT spectra is 15$\arcsec$, and
that of the SMA HC$^{15}$N spectra 1$\farcs$6 $\times$ 1$\farcs$3 (P.A. =
14\degr).  The rms noise level of the JCMT, SMA, and the combined spectra
is 0.13 K, 7.4 K, and 5.6 K, respectively.
\label{fig2}}
\end{figure}

\clearpage
\begin{figure}
\epsscale{.99}
\plotone{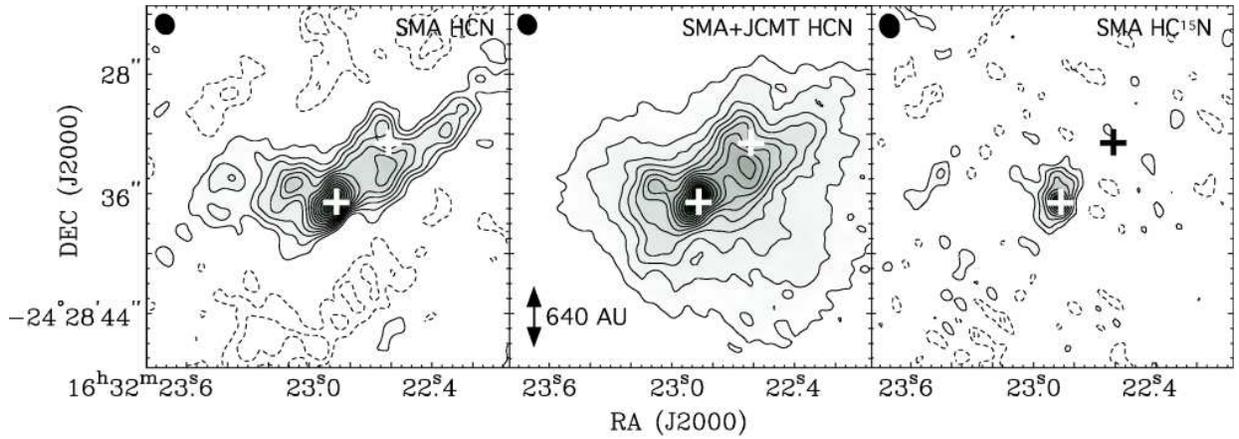}
\caption{Total integrated intensity maps in I16293 taken with the SMA
(left) and SMA+JCMT (middle) in the HCN (4--3) emission, and the HC$^{15}$N
(4--3) emission with the SMA (right). Crosses in each panel indicate
positions of the 354.5 GHz continuum emission observed with the SMA (Fig.1,
Table 2).  Filled ellipses at the top left corner of each panel show the
synthesized beam, that is, 1$\farcs$3 $\times$ 1$\farcs$2 (P.A. = 30\degr)
in the SMA and the combined map, and 1$\farcs$6 $\times$ 1$\farcs$3 (P.A. = 14\degr)
in the SMA HC$^{15}$N map.  Contour levels are from 33.50 K km s$^{-1}$ in
steps of 22.33 K km s$^{-1}$ in the SMA and combined HCN map (3$\sigma$ in
steps of 2$\sigma$ in the SMA HCN map), and from 3$\sigma$ in steps of
2$\sigma$ (1$\sigma$ = 6.49 K km s$^{-1}$) in the SMA HC$^{15}$N map.
\label{fig3}}
\end{figure}

\clearpage
\begin{figure}
\includegraphics[angle=-90,scale=3.7]{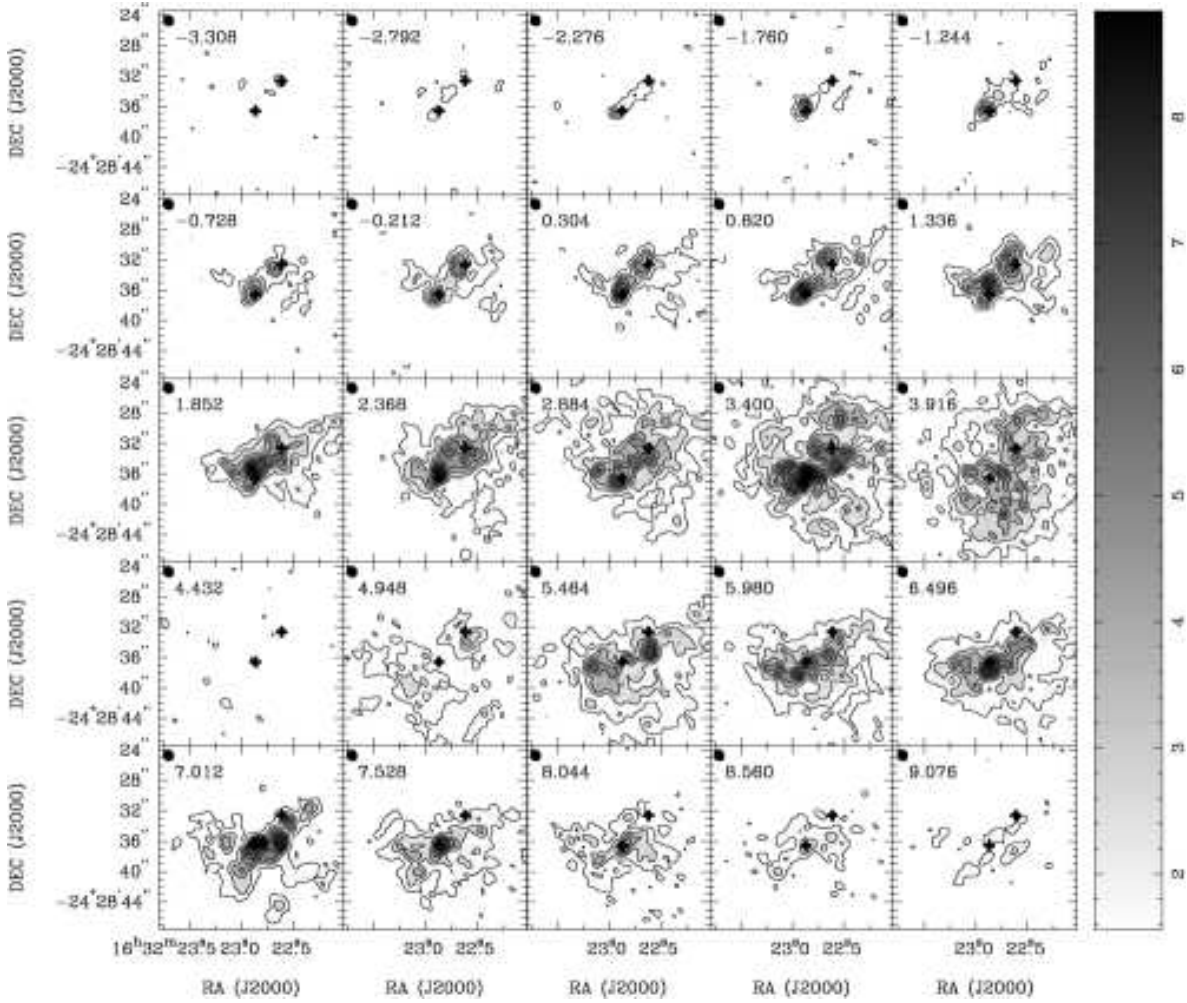}
\caption{SMA + JCMT velocity channel maps of I16293 in the HCN (4--3) line.
Each panel is an average of 3 channels (1 channel = 0.172 km s$^{-1}$).
Crosses in each panel indicate positions of the 354.5 GHz continuum
emission observed with the SMA (Fig.1, Table 2). The synthesized beam size
is shown as a filled ellipse at the top left corner of each panel.  Numbers
at the upper-left corner of each panel are $V_{LSR}$ (km s$^{-1}$).
Contour levels are from 2$\sigma$ in steps of 2$\sigma$ (1$\sigma$ = 3.23
K).
\label{fig4}}
\end{figure}

\clearpage
\begin{figure}
\includegraphics[angle=0,scale=3.7]{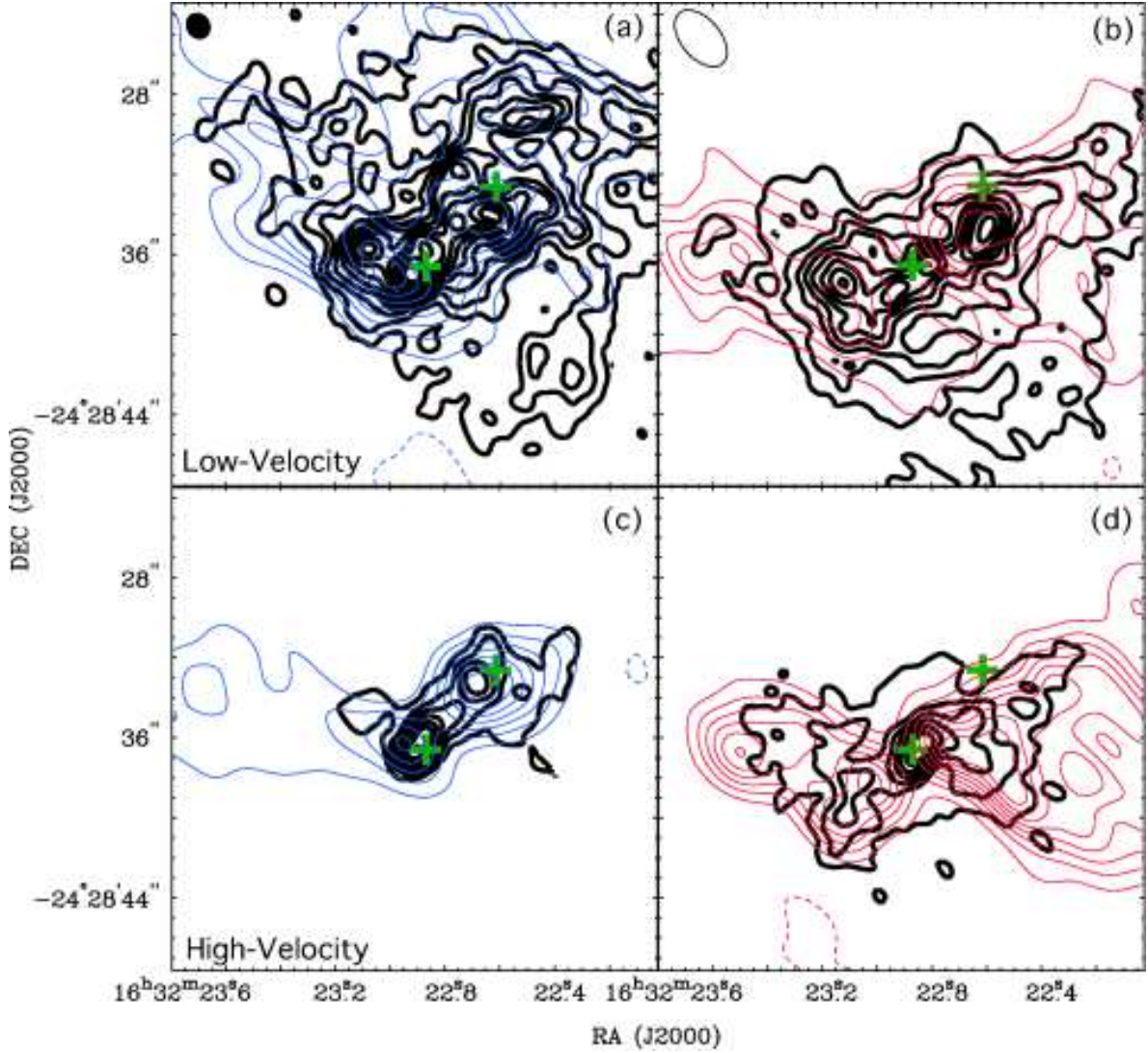}
\caption{SMA + JCMT velocity channel maps of I16293 in the HCN (4--3)
emission at four different velocity ranges: (a) low-velocity blueshifted
range (2.54 - 3.40 km s$^{-1}$); (b) low-velocity redshifted range (4.95 -
6.50 km s$^{-1}$); (c) high-velocity blueshifted range (-2.62 - +0.65 km
s$^{-1}$); and (d) high-velocity redshifted range (7.36 - 9.08 km
s$^{-1}$), superposed on the CO (2--1) outflow map taken with the SMA (blue
and red contours; Yeh et al. 2007).  Contour levels are from 5.30 (K) in
steps of 3.54 (K) in the HCN map and 5.68 (K) in steps of 5.68 (K) in the
CO map.  Green crosses indicate positions of the protobinary (Fig.1, Table
2).  A filled ellipse at the top left corner of panel (a) indicates the
synthesized beam of the HCN observations (1$\farcs$3 $\times$ 1$\farcs$2;
P.A. = 30\degr), while an open ellipse at the top left corner of panel (b)
the synthesized beam size of the CO map (3$\farcs$3 $\times$ 2$\farcs$0;
P.A. = 40\degr).
\label{fig5}}
\end{figure}

\clearpage
\begin{figure}
\includegraphics[angle=0,scale=.85]{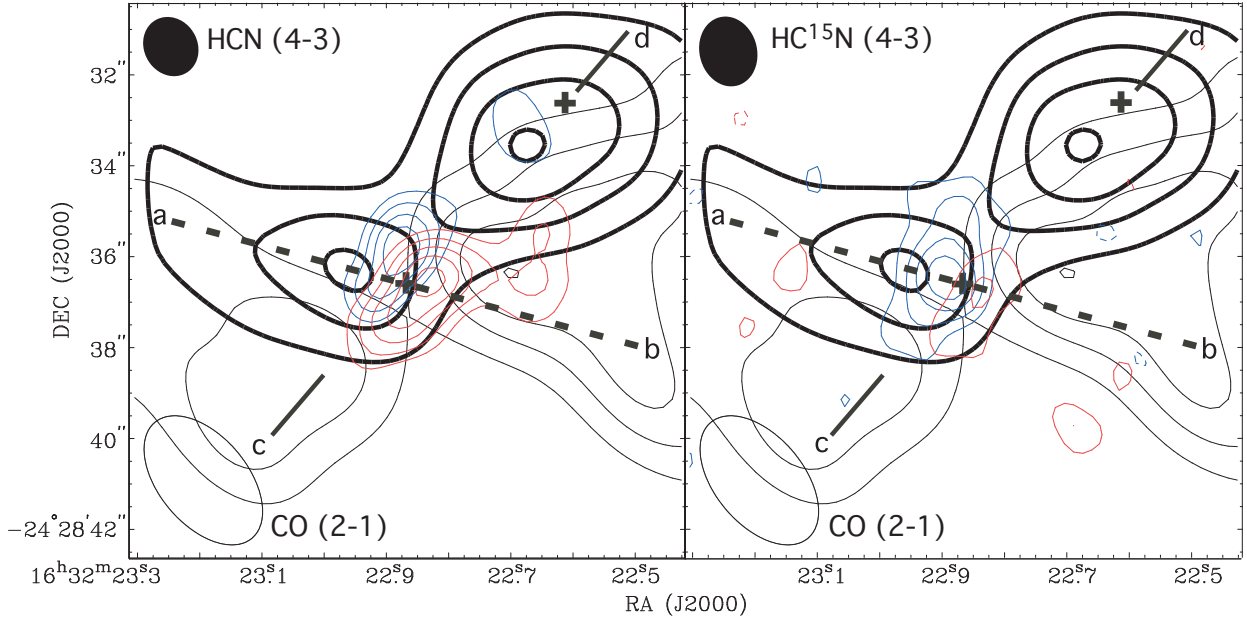}
\caption{High-velocity blueshifted ($V_{LSR}$ = -2.62 -- 2.37 km s$^{-1}$)
and redshifted ($V_{LSR}$ = 5.98 -- 9.08 km s$^{-1}$) HCN (4--3) emission
taken with the SMA and JCMT (left) and HC$^{15}$N (4--3) emission taken
with the SMA (right) in I16293. The velocity range is set to be wider than
that of Figure 5 in order to improve the signal-to-noise ratio of the
HC$^{15}$N image.  The CO (2--1) outflow map taken with the SMA
(Yeh et al. 2007) is overlaid
on the HCN and HC$^{15}$N images. The blueshifted and redshifted components
in the HCN and HC$^{15}$N lines are shown in blue and red contours, while
the blueshifted and redshifted CO outflow in black bold and thin
contours, respectively. Contour levels are from 19.4 (K) in steps of 3.5
(K) in the HCN map, from 3.2 (K) in steps of 2.1 (K) in the HC$^{15}$N map,
and from 17.1 K in steps of 11.4 K in the CO map.  Crosses indicate
positions of the 354.5 GHz continuum emission observed with the SMA (Fig.1,
Table 2). A filled ellipse at the top left corner shows the synthesized
beam of the HCN (1$\farcs$3 $\times$ 1$\farcs$2; P.A. = 30\degr) and
HC$^{15}$N (1$\farcs$6 $\times$ 1$\farcs$3; P.A. = 14\degr) images, while
open ellipses at the bottom left corner the synthesized beam of the CO
(2--1) image (3$\farcs$3 $\times$ 2$\farcs$0; P.A. = 40\degr).  Lines a-b
and c-d indicate cuts of the Position-Velocity diagrams shown in
Figure 7 and 8, respectively.
\label{fig6}}
\end{figure}

\clearpage
\begin{figure}
\includegraphics[angle=0,scale=.75]{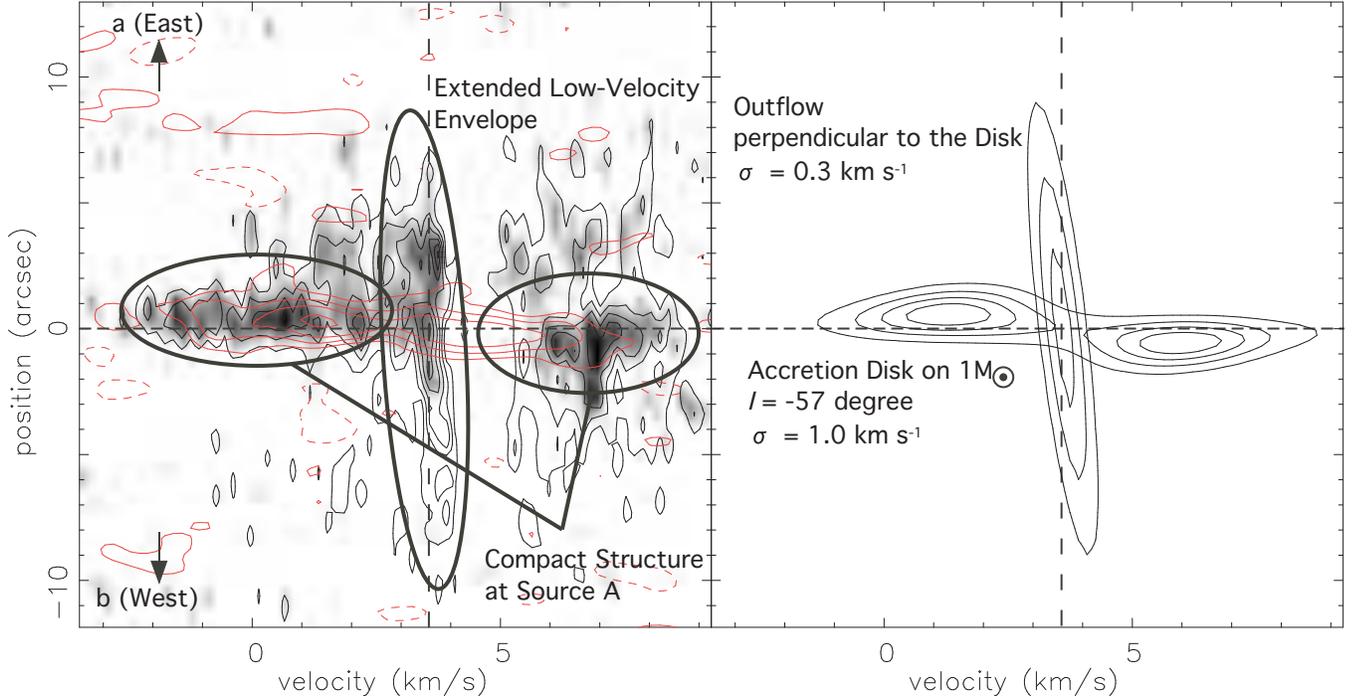}
\caption{Left: Position-Velocity (P-V) diagrams along the minor axis of the
compact flattened structure at Source A in the HC$^{15}$N
(4--3) emission (red contour; P.A. = 74\degr), as well as in the HCN (4--3) line with the
SMA + JCMT (black contour) and SMA (gray).
Contour levels are from 11.2 (K) in steps of 11.2 (K) in the HCN emission
and from 4.23 (K) in steps of 4.23 (K) in the HC$^{15}$N emission.  A
horizontal dashed line indicates the position of Source A, and a vertical
dashed line the systemic velocity ($\sim$ 3.6 km s$^{-1}$). There are two
distinct components, one is the high-velocity compact structure seen in the
SMA HCN and HC$^{15}$N images and the other the extended low-velocity
envelope seen in the combined image.  Right: Model P-V diagrams to
interpret the observational P-V diagrams in the left panel. Contour levels
are the same as in the combined HCN P-V diagram. The P-V diagram of the
high-velocity compact structure can be interpreted as an accretion motion
onto the central mass of 1 M$_{\odot}$ in the geometrically thin disk with
an inclination angle of 57$\degr$ from the plan of the sky and an internal
velocity dispersion ($\equiv$ $\sigma$) of 1.0 km s$^{-1}$.  The P-V
diagram of the extended component can be understood as an outflowing gas
perpendicular to the flattened structure with an internal velocity
dispersion of 0.3 km s$^{-1}$. The model emission is assumed to have
Gaussian intensity distribution. The strong self-absorption around
$V_{LSR}$ = 4 - 5 km s$^{-1}$ seen in the real data can not be reproduced
with this ``toy'' model.
\label{fig7}}
\end{figure}

\clearpage
\begin{figure}
\includegraphics[angle=0,scale=.95]{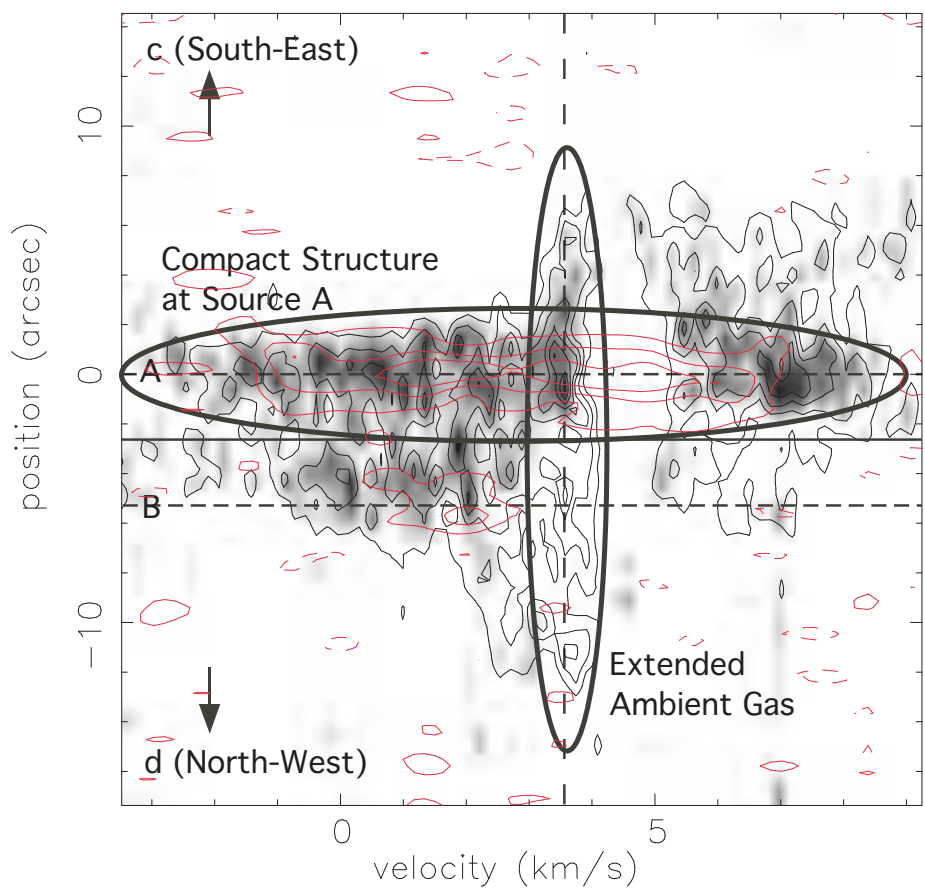}
\caption{P-V diagrams in the HCN (4--3) line with the
SMA + JCMT (black contour) and SMA (gray) and in the HC$^{15}$N (4--3) line
with the SMA (red contour), along the axis joining sources A and B (P.A. = 139\degr).
Contour levels are from 11.2 (K) in steps of 11.2 (K) in the HCN emission
and from 4.23 (K) in steps of 4.23 (K) in the HC$^{15}$N emission.  Upper
and lower horizontal dashed lines show the position of Source A and Source
B, respectively, and a vertical dashed line the systemic velocity ($\sim$
3.6 km s$^{-1}$).  A solid horizontal line indicates the middle point
between Source A and Source B, and should be the binary axis on the
assumption of the same mass in Source A and Source B. 
\label{fig8}}
\end{figure}

\clearpage
\begin{figure}
\epsscale{.90}
\plotone{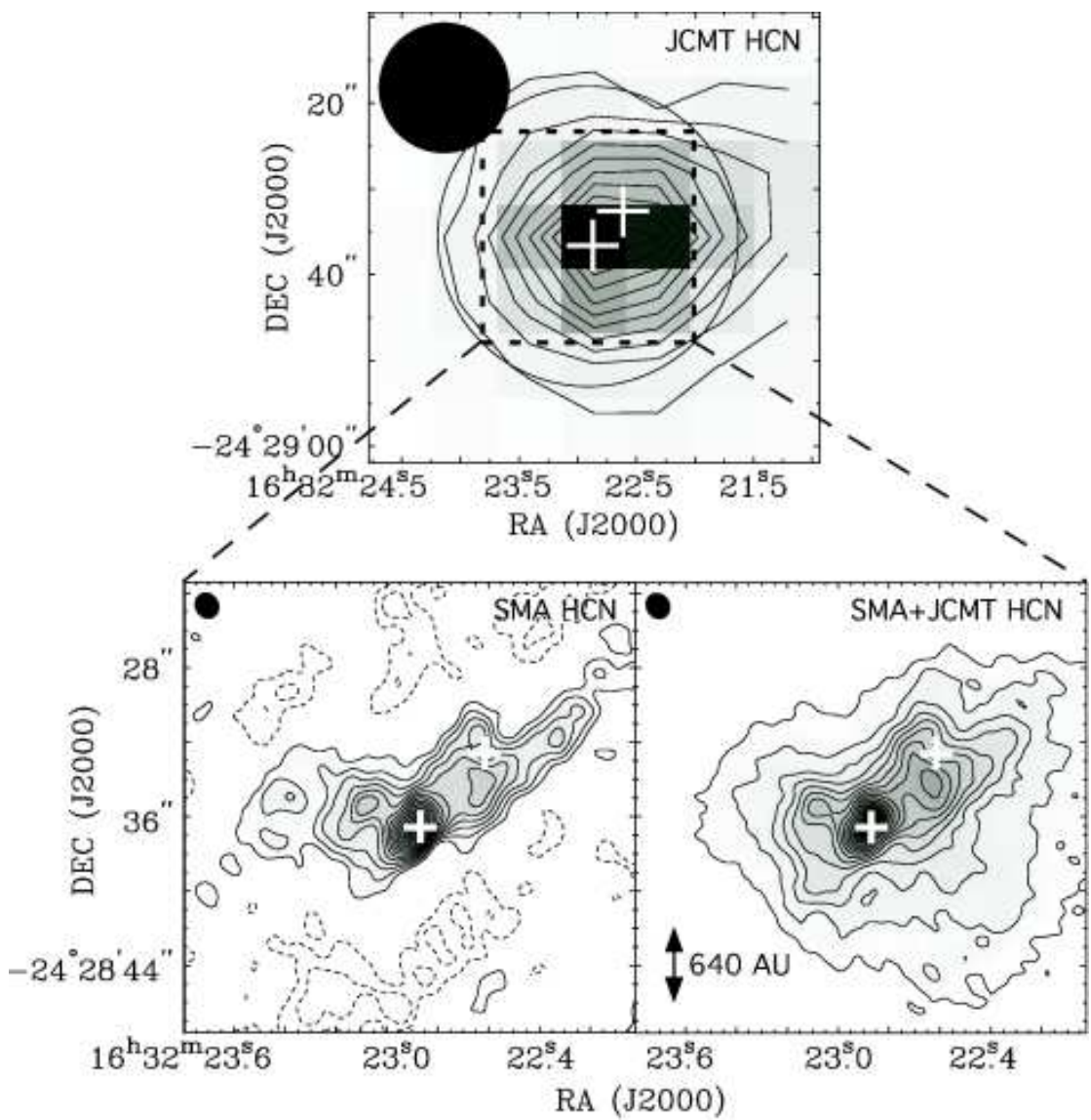}
\caption{Total integrated intensity maps in I16293 taken with JCMT (upper),
the SMA (lower-left), and SMA+JCMT (lower-right) in the HCN (4--3)
emission. An open circle in the JCMT map indicates the field of view of the
SMA, and crosses in each panel indicate positions of the protobinary
(Fig.1, Table 2). Filled ellipses at the top left corner of each panel show
the synthesized beam, that is, 15$\arcsec$ in the JCMT map and 1$\farcs$3
$\times$ 1$\farcs$2 (P.A. = 30\degr) in the SMA and the combined map.
Contour levels are from 4.76 K km
s$^{-1}$ in steps of 4.76 K km s$^{-1}$ in the JCMT map, and from 33.50 K
km s$^{-1}$ in steps of 22.33 K km s$^{-1}$ in the SMA and combined map. 
\label{fig9}}
\end{figure}

\clearpage
\begin{deluxetable}{lccc}
\tabletypesize{\scriptsize}
\tablecaption{Parameters for the SMA Observations of IRAS 16293-2422
\label{tbl-1}}
\tablewidth{0pt}
\tablehead{\colhead{Parameter} & \multicolumn{3}{c}{Value}\\
\cline{2-4}
\colhead{} & \colhead{2003 March 14} & \colhead{2003 July 12} & \colhead{2004 June 19} }
\startdata
Number of Antennas& 5&  4& 5 \\
\ \ \ Right Ascension (J2000)
   & \multicolumn{3}{c}{16$^{\rm h}$ 32$^{\rm m}$ 22$^{\rm s}$.91}\\
\ \ \ Declination (J2000)
   & \multicolumn{3}{c}{-24\degr 28$\arcmin$ 35\farcs52}\\
Primary Beam HPBW& \multicolumn{3}{c}{$\sim$ 35$\arcsec$}\\
Synthesized Beam HPBW (HCN)&
  \multicolumn{3}{c}{1\farcs3 $\times$ 1\farcs2 (P.A. = 30\degr)}\\
Synthesized Beam HPBW (HC$^{15}$N)&
  \multicolumn{3}{c}{1\farcs6 $\times$ 1\farcs3 (P.A. = 14\degr)}\\
Synthesized Beam HPBW (Continuum; uniform)&
  \multicolumn{3}{c}{1\farcs1 $\times$ 0\farcs6 (P.A. = 39\degr)}\\
Baseline Coverage& \multicolumn{3}{c}{10 - 223 (k$\lambda$)}\\
Conversion Factor (Line) & \multicolumn{3}{c}{1 (Jy beam$^{-1}$) = 6.2 (K)}\\
Frequency Resolution (HCN) & \multicolumn{3}{c}{203.125 kHz $\sim$ 0.172 km s$^{-1}$}\\
Frequency Resolution (HC$^{15}$N) & \multicolumn{3}{c}{812.5 kHz $\sim$ 0.708 km s$^{-1}$}\\
Bandwidth&82 MHz $\times$ 8 &82 MHz $\times$ 12 &82 MHz $\times$ 24    \\
Gain Calibrator &nrao530     &1743-038    &  1743-038 \\
Flux of the Gain Calibrator &1.5 Jy & 1.5 Jy & 2.3 Jy \\
Passband Calibrator &Mars       &Mars, Uranus   &Jupiter, Uranus\\
                    &           &               &3c279, 1924-292 \\
Flux Calibrator     &Callisto   &Uranus         &Uranus                           \\
System Temperature (DSB) & \multicolumn{3}{c}{$\sim$ 350 K}\\
rms noise level (Continuum)& \multicolumn{3}{c}{$\sim$ 0.06 Jy beam$^{-1}$}\\
rms noise level (Line; SMA + JCMT)& \multicolumn{3}{c}{$\sim$ 0.9 Jy beam$^{-1}$ / 203.125 kHz}\\
\enddata
\end{deluxetable}

\clearpage

\begin{deluxetable}{lccccc}
\tabletypesize{\scriptsize}
\tablecaption{354.5 GHz Continuum Results in I16293 with the SMA \label{tbl-2}}
\tablewidth{0pt}
\tablehead{\colhead{Protostar} & \colhead{R.A.\tablenotemark{a}} &\colhead{Decl.\tablenotemark{a}} 
&\colhead{Deconvolved FWHM Size\tablenotemark{a}} &\colhead{P.A.\tablenotemark{a}}
&\colhead{Total Flux\tablenotemark{a}} \\
\colhead{} &\colhead{(J2000)} &\colhead{(J2000)} &\colhead{(AU $\times$ AU)} &\colhead
{(\degr)} &\colhead{(mJy)} }
\startdata
Source A &16$^{\rm h}$32$^{\rm m}$22$^{\rm s}$.87 &-24\degr28$\arcmin$36\farcs6 &300 $\times$ 
150  &32.6  &3840  \\
Source B &16$^{\rm h}$32$^{\rm m}$22$^{\rm s}$.62 &-24\degr28$\arcmin$32\farcs6 &150 $\times$ 
140  &24.8  &4050  \\
\enddata
\tablenotetext{a}{Estimated from 2-dimensional Gaussian fittings to the image.}
\end{deluxetable}

\end{document}